\documentclass[twocolumn,english,aps,prb,floats,showpacs]{revtex4}
\usepackage[T1]{fontenc}
\usepackage[latin9]{inputenc}
\usepackage{babel}
\usepackage{bm}
\usepackage{amsmath}
\usepackage{amssymb}
\usepackage{graphicx}
\usepackage{wasysym}
\usepackage{esint}
\usepackage[unicode=true,
 bookmarks=false,
 breaklinks=false,pdfborder={0 0 1},backref=section,colorlinks=false]
 {hyperref}
\usepackage{breakurl}

\makeatletter
\@ifundefined{textcolor}{}
{%
 \definecolor{BLACK}{gray}{0}
 \definecolor{WHITE}{gray}{1}
 \definecolor{RED}{rgb}{1,0,0}
 \definecolor{GREEN}{rgb}{0,1,0}
 \definecolor{BLUE}{rgb}{0,0,1}
 \definecolor{CYAN}{cmyk}{1,0,0,0}
 \definecolor{MAGENTA}{cmyk}{0,1,0,0}
 \definecolor{YELLOW}{cmyk}{0,0,1,0}
}

\@ifundefined{textcolor}{}{%
 \definecolor{BLACK}{gray}{0}
 \definecolor{WHITE}{gray}{1}
 \definecolor{RED}{rgb}{1,0,0}
 \definecolor{GREEN}{rgb}{0,1,0}
 \definecolor{BLUE}{rgb}{0,0,1}
 \definecolor{CYAN}{cmyk}{1,0,0,0}
 \definecolor{MAGENTA}{cmyk}{0,1,0,0}
 \definecolor{YELLOW}{cmyk}{0,0,1,0}
}

\usepackage{babel}
\usepackage{ifpdf}\usepackage{bm}

\makeatother

\begin{document}

\title{Biskyrmions Lattices in Centrosymmetric Magnetic Films}

\author{Daniel Capic, Dmitry A. Garanin, and Eugene M. Chudnovsky}

\affiliation{Physics Department, Herbert H. Lehman College and Graduate School,
The City University of New York, 250 Bedford Park Boulevard West,
Bronx, New York 10468-1589, USA }

\date{\today}
\begin{abstract}
Theoretical framework is developed that permits construction of biskyrmion lattices
observed in non-chiral magnetic films. We study films of finite thickness
containing up to $1000\times1000\times100$ spins. Hexatic biskyrmion
lattices in a pure 2D exchange model are naturally described by the
Weierstrass $\wp$ and $\zeta$ elliptic functions. Starting with
such a lattice as an initial state we investigate how it evolves towards
a minimum-energy state in a zero magnetic field in the presence of perpendicular magnetic
anisotropy (PMA) and dipole-dipole interaction. Metastable biskyrmion lattices exist at low PMA. 
At higher PMA we observe stable triangular lattices of biskyrmion bubbles containing Bloch lines, 
whose energies are lower than the energy of the uniformly magnetized state. 
\end{abstract}

\maketitle

\section{Introduction}

Skyrmions were initially introduced in nuclear physics as solutions
of the nonlinear $\sigma$-model that can describe atomic nuclei \cite{SkyrmePRC58,Polyakov-book,Manton-book}.
They possess topological charge: $Q=\pm1,\pm2$, etc. Different $Q$
arise from different homotopy classes of the mapping of a three-component
fixed-length field onto the $xy$ plane. $Q=1$ would correspond to
a single nucleon, $Q=2$ could describe a deuteron \cite{D1,D2},
$Q=4$ would provide a model of an $\alpha$-particle, etc.

In magnetic films skyrmions are defects of the ferromagnetic order.
They represent a very active field of research due to their nanoscale
size and potential for a dense topologically protected data storage
and information processing \cite{Nagaosa2013,Zhang2015,Klaui2016,Leonov-NJP2016,Hoffmann-PhysRep2017,Fert-Nature2017}.
Much bigger micron-size magnetic bubbles intensively studied in 1970s
\cite{MS-bubbles,ODell} possessed similar topological properties.
They were cylindrical domains surrounded by narrow domain walls. On
the contrary, a typical skyrmion would be small compared to the domain
wall thickness, making it conceptually similar to the topological
objects studied in nuclear physics \cite{BelPolJETP75,Lectures}.

Research on magnetic skyrmions focuses on their stability and dynamics.
Magnetic bubble lattices are in effect domain structures \cite{MS-bubbles,ODell,Ezawa-PRL2010,Makhfudz-PRL2012}
stabilized by the perpendicular magnetic anisotropy (PMA), dipole-dipole
interaction (DDI), and the external magnetic field. On the contrary,
in the absence of other interactions, small skyrmions collapse \cite{CCG-PRB2012}
due to violation of the scale invariance of the 2D exchange model
by the atomic lattice. They can be stabilized by Dzyaloshinskii-Moriya
interaction that is present in materials lacking inversion symmetry
(DMI) \cite{Bogdanov-Nature2006,Heinze-Nature2011,Boulle-NatNano2016,Leonov-NJP2016},
or by quenched randomness \cite{EC-DG-PRL2018,EC-DG-NJP2018} that
has effect similar to the DMI.

\begin{figure}[ht]
\centering{}\includegraphics[width=8cm]{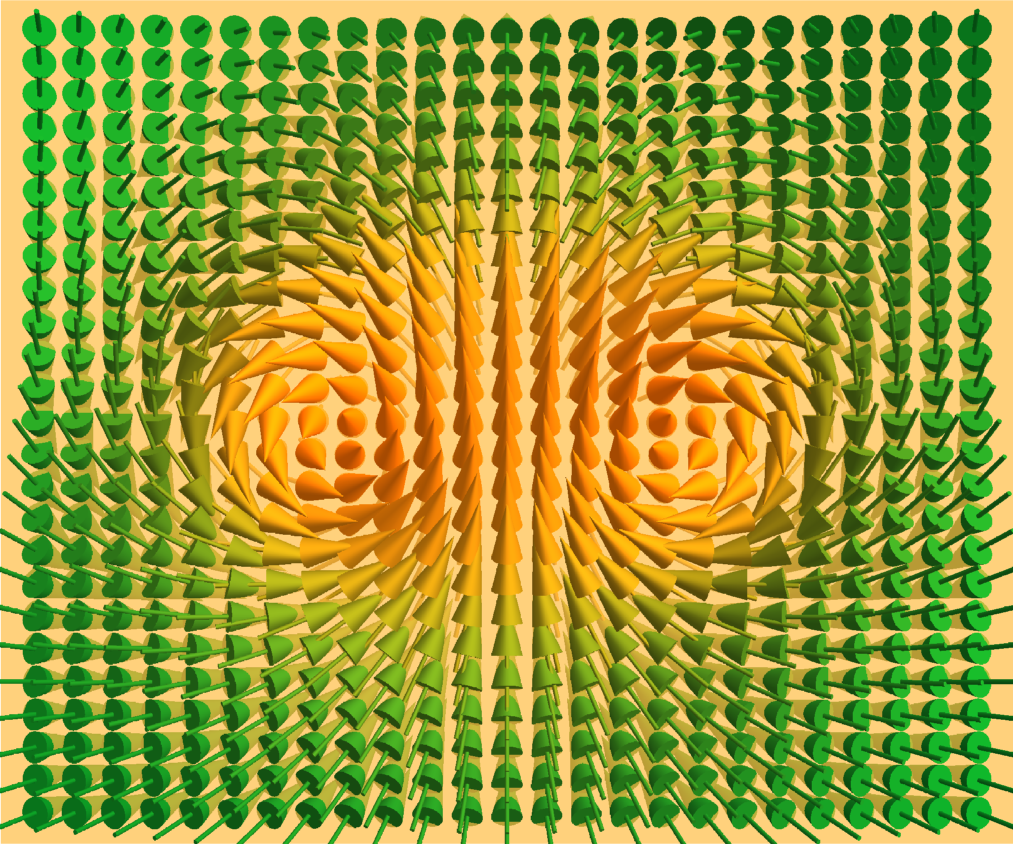} \caption{Computer generated spin field in a Bloch-type byskirmion in a ferromagnetic
film.}
\label{Fig-biskyrmion} 
\end{figure}

While $Q=1$ skyrmions and lattices of such skyrmions have been intensively
studied, observation of Q = 2 biskyrmions, see Fig.\ \ref{Fig-biskyrmion},
and biskyrmion lattices has been rare. Unlike skyrmions, that have
been mostly observed in chiral films, stable biskyrmions lattices have been initially reported down to zero field 
in two non-chiral films of sufficient thickness: the La$_{2-2x}$Sr$_{1+2x}$Mn$_{2}$O$_{7}$
manganite \cite{Yu-2014} and the (Mn$_{1-x}$Ni$_{x}$)$_{65}$Ga$_{35}$
half Heusler alloy \cite{Zhang2016}. Recent works \cite{Loudon2019,Yao2019,Gobel2019} provided 
further experimental and numerical evidence of stable biskyrmions in centrosymmetric magnetic materials.

Theoretical works on biskyrmions have been scarce. It has been shown \cite{GCZ-EPL2017} that
topological defects with $Q>1$ appear naturally in centrosymmetric
films due to the presence of Bloch lines in labyrinth domains on increasing
magnetic field. Numerical investigation of biskyrmions, including
the current-induced dynamics, was performed in Ref.\ \onlinecite{XZhang2017}
within a 2D frustrated micromagnetic model. Biskyrmions arising from
frustrated Heisenberg exchange have also been reported in the studies
of triangular spin lattices \cite{Leonov-NatCom2015} and in a model
based upon Ginzburg-Landau theory \cite{Lin-PRB2016}. Metastable
biskyrmion configurations have been observed in Landau-Lifshitz dynamics
of a frustrated bilayer film \cite{Nowak2017}. However, the study
of biskyrmion lattices has been absent so far.

In real systems, the magnetic biskyrmions are more complicated than
generic $Q=2$ solutions of the Belavin-Polyakov (BP) 2D exchange
model. They are formed by a number of competing interactions on top
of the exchange, such as PMA and DDI. The latter rules out any meaningful
analytical solution. In this paper we adopt the following approach.
First, we prepare a biskyrmion lattice that is a solution of the pure
exchange model. Fortunately, it can be described in the complex plane,
$z=x+iy$, by standard elliptic Weierstrass functions that have been
previously used to build multiskyrmion configurations in nuclear physics
\cite{Manton-book}. We then turn on the PMA and DDI and compute numerically
evolution of the system towards the spin configuration that corresponds
to the energy minimum. 

Our calculations are performed on 3D lattices
containing $1000\times1000\times100$ spins that describe magnetic
films of finite thickness. In accordance with real experiments, we find
that the lowest energy achieved in our numerical experiments corresponds to a triangular lattice of biskyrmion bubbles
containing Bloch lines.

The paper is organized as follows. Analytical expressions for individual
biskyrmion and biskyrmion lattices within the BP model are given in
Section \ref{Sec_BP}. Biskyrmion lattices in a 2D magnetic film with
the PMA and DDI are studied numerically within a discrete spin model
in Section \ref{Sec_Film}. Our results are summarized in Section \ref{Sec_Discussion}.

\section{Biskyrmion lattices in a 2D exchange model}

\label{Sec_BP}

The 2D exchange model is described by the energy 
\begin{equation}
{\cal H}_{ex}=\frac{J}{2}\int dxdy\left(\frac{\partial \bold{s}}{\partial x} \cdot \frac{\partial \bold{s}}{\partial x} + \frac{\partial \bold{s}}{\partial y} \cdot \frac{\partial \bold{s}}{\partial y}\right),\label{Energy_exchange_continuous}
\end{equation}
where ${\bf s}$ is a three-component fixed-length spin field, ${\bf s}^{2}=1$,
$J$ is the exchange constant, and summation over spin components
$\alpha=x,y,z$ is performed. We choose uniform magnetization, ${\bf s}=(0,0,-1)$,
in the negative z-direction at infinity. Spin-field configurations
belong to homotopy classes characterized by the topological charge
\begin{equation}
Q=\int\frac{dxdy}{4\pi}\:{\bf s}\cdot\frac{\partial{\bf s}}{\partial x}\times\frac{\partial{\bf s}}{\partial y}\label{Q}
\end{equation}
that takes quantized values $Q=0,\pm1,\pm2,...$. The value of $Q$
shows how many times the spin vector circumscribes the full body angle
$4\pi$ as the position vector covers the whole $xy$ plane. 

In each homotopy class the energy is minimized by ${\bf s}(x,y)$ satisfying
\begin{equation}
\mathbf{s}\times\bm{\nabla}^{2}\mathbf{s}=0\label{equilibrium},
\end{equation}
which means that the spins are collinear with the exchange field. In
2D this is equivalent to
\begin{equation}
\frac{\partial\mathbf{s}}{\partial x}=\pm\mathbf{s}\times\frac{\partial\mathbf{s}}{\partial y},\qquad\frac{\partial\mathbf{s}}{\partial y}=\mp\mathbf{s}\times\frac{\partial\mathbf{s}}{\partial x}.\label{Equil_simpler_vector_xy}
\end{equation}
One recovers Eq. (\ref{equilibrium}) by differentiating the first of these equations over $y$, the second one over $x$, and subtracting one from the other.
With the help of Eqs.\ (\ref{Energy_exchange_continuous}) and (\ref{Equil_simpler_vector_xy}) it is easy to obtain
the relation between the exchange energy and the topological charge \cite{BelPolJETP75}:
\begin{equation}
E=4\pi J|Q|.\label{Energy_Q}
\end{equation}

Extremal spin configurations can be obtained by mapping the problem onto the complex plane, $z=x+iy$. In terms
of the complex function 
\begin{equation}
\omega(z) = \frac{s_{x}+is_{y}}{1-s_{z}}\label{omega_via_s}
\end{equation}
Eqs. (\ref{Equil_simpler_vector_xy}) reduce to linear equations,
\begin{equation}
\frac{\partial\omega}{\partial x}=\pm i\frac{\partial\omega}{\partial y},\label{CR_condition}
\end{equation}
that are familiar Cauchy-Riemann (CR) conditions of the analyticity of the function. The corresponding spin configuration follow from the relations
\begin{equation}
s_{x}+is_{y}=\frac{2\omega}{|\omega|^{2}+1},\quad s_{z}=\frac{|\omega|^{2}-1}{|\omega|^{2}+1}.
\end{equation}
The remarkable property of the model is that any analytic function $\omega(z)$ provides solution for the extremal spin configuration. 

In particular, the sum of poles
\begin{equation}
\omega(z)=\sum_{i}\frac{\lambda_{i}e^{i\gamma_{i}}}{z-z_{i}},\label{omega_sum-of-poles-antiskyrmions}
\end{equation}
describes a collection of antiskyrmions with spins at the poles pointing up against the spin-down
background, the parameters $z_{i}$, $\lambda_{i}$, $\gamma_{i}$
defining the position, size, and chirality angle of the $i$-th
antiskyrmion. This function satisfies the CR condition, Eq. (\ref{CR_condition}),
with the plus sign. In-plane spin components are rotating counterclockwise
as the observation point is moving clockwise around the center (the pole) of
the antiskyrmion. Each antiskyrmion is contributing -1
to the total topological charge $Q$. 

The state with the skyrmions whose central spins are pointing
down against the spin-up background can be obtained by rotating all
spins of the above collection of antiskyrmions by the angle
$\pi$ around the $x$-axis. Conformal invariance of the 2D exchange model allows one to achieve the same effect
by taking the reciprocal of the $\omega$-function, 
\begin{equation}
\omega(z)=\left(\sum_{i}\frac{\lambda_{i}e^{i\gamma_{i}}}{z-z_{i}}\right)^{-1}. \label{omega_sum-of-poles-skyrmions-down}
\end{equation}
It describes a collection of skyrmions pointing down against the spin-up background, with each skyrmion corresponding
to a zero of $\omega(z)$. Since the rotation of all spins preserves the homotopy class, such skyrmions have topological charge
$Q=-1$, with in-plane spin components rotating clockwise when the observation point moves clockwise around the skyrmion's center. This illustrates a less appreciated fact that the sign of $Q$ is determined by both, the topology of the solution and the boundary condition (direction of spins at infinity). 

The expression with a complex-conjugated argument
\begin{equation}
\omega(z)=\sum_{i}\frac{\lambda_{i}e^{i\gamma_{i}}}{z^{*}-z_{i}^{*}},\label{omega_sum-of-poles-skyrmions}
\end{equation}
that satisfies the CR condition Eq. (\ref{CR_condition}) with the minus  sign, describes a collection of skyrmions whose central spins are pointing up against the spin-down background. Here, each skyrmion has topological charge $Q=1$.

The energy of the collection of skyrmions or collection of antiskyrmions is entirely defined by the total topological charge $Q$ through Eq.\ (\ref{Energy_Q}). Q determined by the number of poles and is independent of $z_{i}$, $\lambda_{i}$, and $\gamma_{i}$ due to the symmetry of the 2D exchange model. On the contrary, any combination of skyrmions and antiskyrmions yields a non-analytical $\omega(z)$ that violates the CR conditions and, hence, does not satisfy Eq. (\ref{equilibrium}). In the physical language, this means
that such a configuration cannot be static. Skyrmions and antiskyrmions must annihilate while conserving the total topological charge $Q$, with the excess energy going into spin waves. 

For an arbitrary $Q$, a skyrmion solution centered at $z=z_{0}$
is given by 
\begin{equation}
\omega(z)=\left(\frac{\lambda}{z^{*}-z_{0}^{*}}\right)^{Q}e^{i\gamma}.\label{Q_arb_skyrmion}
\end{equation}
The antiskyrmion solution is obtained by replacing $z^{*}$ and $z_{0}^{*}$
with $z$ and $z_{0}$. A biskyrmion with $Q=2$ and separation $d$
between two skyrmions of chirality $\gamma_{1}$ and $\gamma_{2}$
in the biskyrmion, centered at $z=0$, can be written as 
\begin{equation}
\omega(z)=\frac{\lambda e^{i\gamma_{1}}}{z^{*}+d/2}+\frac{\lambda e^{i\gamma_{2}}}{z^{*}-d/2}.\label{biskyrmion}
\end{equation}
A generic Bloch-type biskyrmion with opposite chiralities $\gamma_{1}=-\gamma_{2}=\pi/2$
is shown in Fig.\ \ref{Fig-biskyrmion}.

\begin{figure}[ht]
\centering{}\includegraphics[width=8cm]{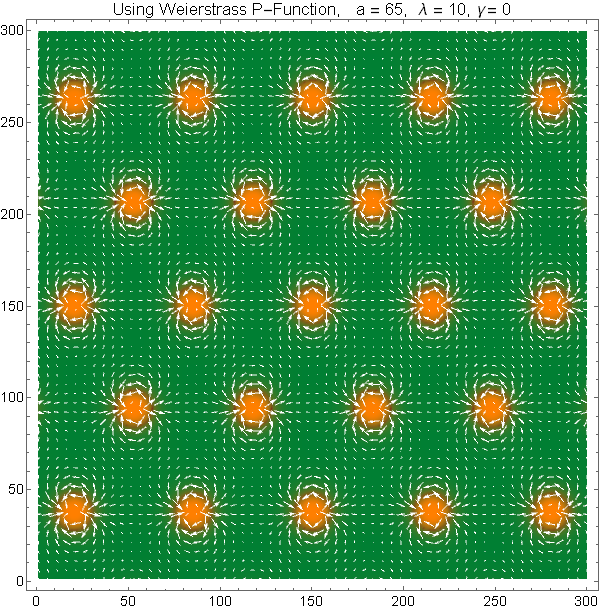} 
\caption{Triangular $Q=2$ skyrmion lattices generated with the help of the
Weierstrass $\wp$-function. }
\label{Fig-wp-lattice} 
\end{figure}

Regular lattices of topological defects (skyrmions or antiskyrmions),
can be constructed using functions that are double-periodic in the complex
plane $z=x+iy$. The elliptic Weierstrass $\wp$-function provides
such a solution for the periodic lattice of $Q=2$ skyrmions, 
\begin{eqnarray}
\wp(z^{*}) & = & \frac{1}{z^{*2}}+\sum_{m,n\neq0}\left[\frac{1}{(z^{*}+2mp_{1}+2np_{2})^{2}}\right.\nonumber \\
 & - & \left.\frac{1}{(2mp_{1}+2np_{2})^{2}}\right].\label{wp}
\end{eqnarray}
Here $m,n$ are integers, and $p_{1,2}$ are half periods of the elliptic
function satisfying ${\rm Im}(p_{2}/p_{1})\neq0$. Consequently $\wp(z^{*})=\wp(z^{*}+2mp_{1})=\wp(z^{*}+2np_{2})$,
so that $Q = 2$ skyrmions are found at $z^{*}=2mp_{1}+2np_{2}$.

Following experimental observations we will be interested in the triangular
biskyrmion lattices. They correspond to the choice 
\begin{equation}
p_{1}=\frac{a'}{2},\qquad p_{2}=\frac{-1+i\sqrt{3}}{4}a',
\end{equation}
where $a'$ is the skyrmion lattice spacing. The choice for the $\omega$-function
is $\omega=\lambda^{2}e^{-i\gamma}\wp(z^{*})$, where $\lambda$ and
$\gamma$ represent the size and chirality of skyrmions in the lattice.
Two such lattices with $\gamma=0$ and $\gamma=\pi/2$ are shown in
Fig.\ \ref{Fig-wp-lattice}.

\begin{figure}[ht]
\centering{}\includegraphics[width=8cm]{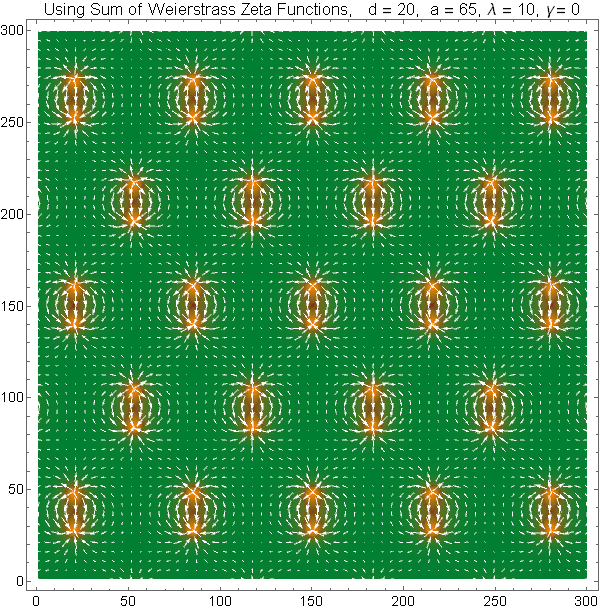} \includegraphics[width=8cm]{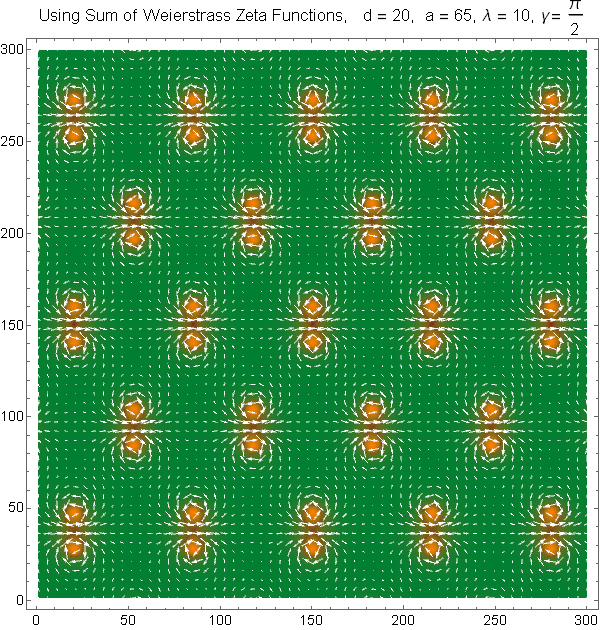}
\caption{Triangular biskyrmion lattices generated with the help of Weierstrass
$\zeta$-functions. Upper panel: N\'eel biskyrmions, $\gamma=0$. Lower
panel: Bloch biskyrmions, $\gamma=\pi/2$.}
\label{Fig-wz-lattice} 
\end{figure}

So far the lattices we have built had zero separation $d$ between
the two skyrmions in the biskyrmion, see Eq.\ (\ref{biskyrmion}).
To build biskyrmion lattices with a finite $d$, one can use Weierstrass
$\zeta$-function defined by $d\zeta(z^{*})/dz^{*}=-\wp(z^{*})$.
Its explicit form is given by 
\begin{eqnarray}
\zeta(z^{*}) & = & \frac{1}{z^{*}}+\sum_{m,n\neq0}\left[\frac{1}{z^{*}-2np_{1}-2mp_{2}}\right.\nonumber \\
 & + & \left.\frac{1}{2np_{1}+2mp_{2}}+\frac{z^{*}}{(2np_{1}+2mp_{2})^{2}}\right].
\end{eqnarray}
This elliptic function is only quasi-periodic, satisfying $\zeta(z^{*}+2mp_{1}+2np_{2})=\zeta(z^{*})+2m\zeta(p_{1})+2n\zeta(p_{2})$.
However, the function 
\begin{equation}
\omega(z^{*})=\frac{\lambda^{2}}{d}e^{i\gamma}\left[\zeta\left(z^{*}+\frac{d}{2}e^{i\eta}\right)-\zeta\left(z^{*}-\frac{d}{2}e^{i\eta}\right)\right]\label{zz}
\end{equation}
is double-periodic. It describes the lattice of biskyrmions of opposite chiralities that have a finite separation $d$ between skyrmions in a biskyrmion. The value of  $\eta$ selects the orientation of the biskyrmion, with $\eta=0$ and $\eta=\pi/2$ corresponding to the horizontal and vertical orientation respectively. Lattices with vertical orientation of N\'{e}el and Bloch biskyrmions are shown in Fig.\ \ref{Fig-wz-lattice}.

It is interesting to notice that standard elliptic functions are better suited for the description of $Q=2$ slyrmion and biskyrmion lattices than $Q = 1$ skyrmion lattices. 

\section{Biskyrmion lattices in a 2D magnetic film}

\label{Sec_Film}

In the numerical work we study a lattice model of a ferromagnetic
film of finite thickness with the energy given by the sum over lattice
sites $i,j$ 
\begin{eqnarray}
{\cal H} & = & -\frac{1}{2}\sum_{ij}J_{ij}\mathbf{s}_{i}\cdot\mathbf{s}_{j}-H\sum_{i}s_{iz}-\frac{D}{2}\sum_{i}s_{iz}^{2} \nonumber \\
&-& \frac{E_{D}}{2}\sum_{ij}\Phi_{ij,\alpha\beta}s_{i\alpha}s_{j\beta}.\label{Hamiltonian}
\end{eqnarray}
Here the exchange coupling is $J$ for the nearest neighbors on a
simple cubic lattice and zero otherwise, $D$ is the easy-axis PMA
constant, $H\equiv g\mu_{B}SB$, with $S$ being the value of
the atomic spin and $B$ being the induction of the applied magnetic
field. In the DDI part of the energy 
\begin{equation}
\Phi_{ij,\alpha\beta}\equiv a^{3}r_{ij}^{-5}\left(3r_{ij,\alpha}r_{ij,\beta}-\delta_{\alpha\beta}r_{ij}^{2}\right),\label{DDI}
\end{equation}
where $\mathbf{r}_{ij}\equiv\mathbf{r}_{i}-\mathbf{r}_{j}$ is the
displacement vector between the lattice sites and $\alpha,\beta=x,y,z$
denote cartesian components in a 3D coordinate space. The parameter $E_{D}=\mu_{0}M_{0}^{2}a^{3}/(4\pi)$
defines the strength of the DDI, with $M_{0}=g\mu_{B}S/a^{3}$ being
the magnetization for our lattice model and $\mu_{0}$ being the magnetic
permeability of vacuum.

The ratio of the PMI and DDI is given by the dimensionless parameter
$\beta\equiv D/(4\pi E_{D})$. For $\beta>1$, the energy of the uniform
state with spins directed along the $z$-axis is lower than that of
the state with spins lying in the film's plane. For $\beta<1$, the
state with spins in the plane has a lower energy. The most interesting
practical case is $\beta\sim1$ realized in many materials due to considerable compensation of the effects of the PMA and
DDI.

An important parameter controlling the DDI is the film thickness represented
by $N_{z}$ in the units of the atomic spacing $a$. For thin films
that are studied here, the magnetization inside the film is nearly
constant along the direction perpendicular to the film. Thus one can
make the problem effectively two-dimensional by introducing the effective
DDI between the columns of parallel spins, considered as effective
spins of the 2D model. This greatly speeds up the computation. To
this end, for the simple cubic lattice, one can write the dipolar
coupling, Eq. (\ref{DDI}), as $\varPhi_{ij,\alpha\beta}=\phi_{\alpha\beta}\left(n_{x},n_{y},n_{z}\right)$,
where $n_{x}\equiv i_{x}-j_{x}$ etc., are the distances on the lattice
and 
\begin{equation}
\phi_{\alpha\beta}\left(n_{x},n_{y},n_{z}\right)=\frac{3n_{\alpha}n_{\beta}-\delta_{\alpha\beta}\left(n_{x}^{2}+n_{y}^{2}+n_{z}^{2}\right)}{\left(n_{x}^{2}+n_{y}^{2}+n_{z}^{2}\right)^{5/2}}.\label{phinnnDef-1}
\end{equation}
The effective DDI is defined by 
\begin{equation}
\bar{\phi}_{\alpha\beta}\left(n_{x},n_{y}\right)=\frac{1}{N_{z}}\sum_{i_{z},j_{z}=1}^{N_{z}}\phi_{\alpha\beta}\left(n_{x},n_{y},i_{z}-j_{z}\right).
\end{equation}
Using the symmetry, one can express this result in the form with only
one summation, 
\begin{eqnarray}
\bar{\phi}_{\alpha\beta}\left(n_{x},n_{y}\right) & = & \phi_{\alpha\beta}\left(n_{x},n_{y},0\right)\\
 & + & \frac{2}{N_{z}}\sum_{n_{z}=1}^{N_{z}-1}\left(N_{z}-n_{z}\right)\phi_{\alpha\beta}\left(n_{x},n_{y},n_{z}\right),\nonumber 
\end{eqnarray}
that is used in the computations. That effective DDI (that can be pre-computed) has different forms in
different ranges of the distance $r$. At $r\apprge aN_{z}$ it scales
as the interaction of magnetic dipoles $1/r^{3}$, while at $r\lesssim aN_{z}$
it goes as $1/r$ that corresponds to the interaction of magnetic
charges at the surface of the film.

\begin{figure}[ht]
\centering{}\includegraphics[width=8cm]{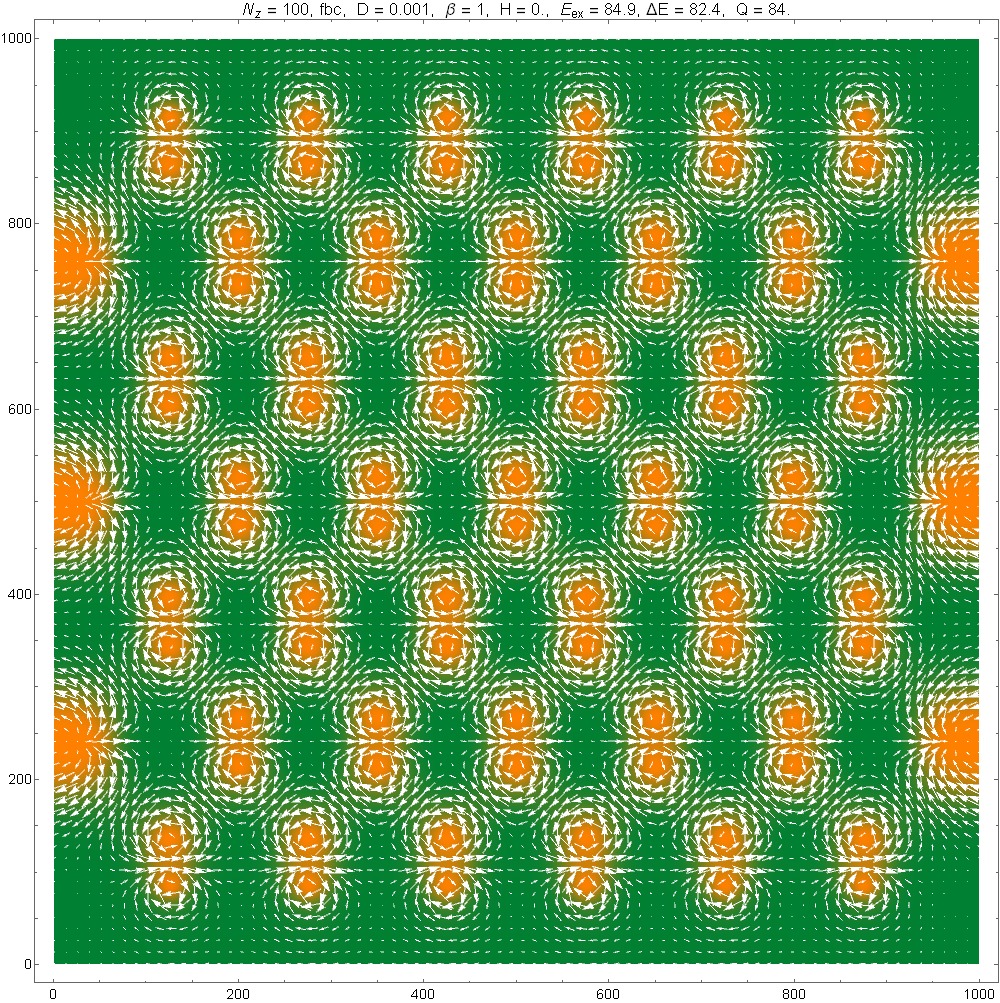} 
\caption{Biskyrmion lattice that evolves from the Weierstrass lattice similar to the one shown in the lower panel of Fig.\ \ref{Fig-wz-lattice}
under the action of PMA and DDI at $H=0$, $\beta=1$, and $D = 0.001$. The energy of this state is higher than the energy of the uniformly magnetized state, $\Delta E = 82.4$. }
\label{Fig-biskyrmion-lattice} 
\end{figure}

\begin{figure}[ht]
\vspace{0.5cm}
\centering{}\includegraphics[width=8cm]{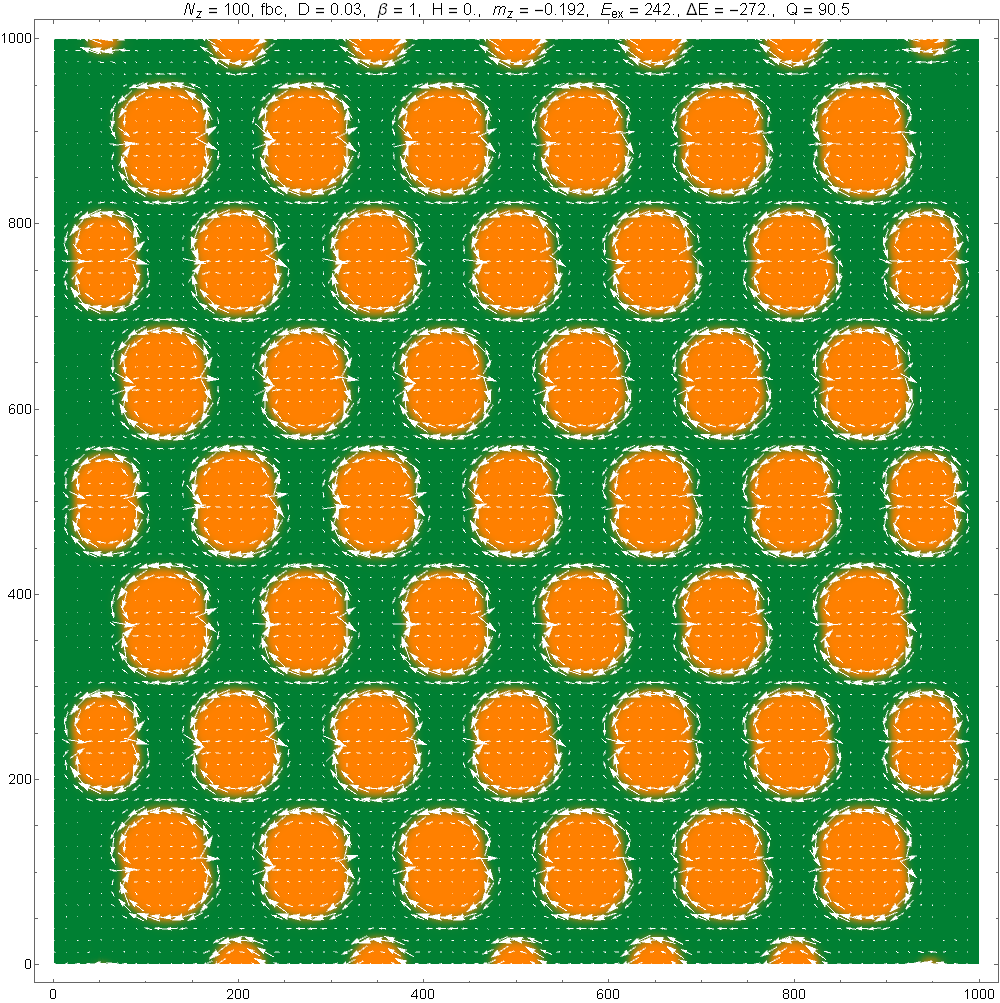}
\caption{Biskyrmion bubble lattice that evolves from the Weierstrass lattice under the action of PMA and DDI at $H=0$, $\beta=1$, and $D = 0.03$. Its energy is lower than the energy of the uniformly magnetized state, $\Delta E = - 272$, indicating that it can be a stable domain structure formed by the DDI. }
\label{Fig-bubble-lattice} 
\end{figure}

Starting with a biskyrmion lattice of the previous section as an initial
condition we compute its evolution towards the minimum-energy configuration
in a system containing up to $1000\times1000\times100$ spins. Our
numerical method \cite{DCP-PRB2013} combines sequential rotations
of spins ${\bf s}_{i}$ towards the direction of the local effective
field, ${\bf H}_{{\rm eff},i}=-\partial{\cal H}/\partial{\bf s}_{i}$,
with the probability $\alpha$, and the energy-conserving spin flips
(so-called \textit{overrelaxation}), ${\bf s}_{i}\to2({\bf s}_{i}\cdot{\bf H}_{{\rm eff},i}){\bf H}_{{\rm eff},i}/H_{{\rm eff},i}^{2}-{\bf s}_{i}$,
with the probability $1-\alpha$. The parameter $\alpha$ plays the
role of the effective relaxation constant. We mainly use the value
$\alpha=0.03$ that provides the overall fastest convergence.

The dipolar part of the effective field takes the longest time to
compute. The most efficient method uses updates
of the dipolar field after all spins are updated rather than after updating each individual spin. Computation of
the dipolar field uses the Fast Fourier Transform (FFT) algorithm
that yields the dipolar fields in the whole sample as one program
step. The total topological charge $Q$ of the lattice is computed
numerically using the lattice-discretized version of Eq.\ (\ref{Q})
in which first derivatives are approximated by the four-point formula.

Computations were performed with Wolfram Mathematica using compilation.
Most of the numerical work has been done on the 20-core Dell Precision
T7610 Workstation. The FFT for computing the DDI was performed via
Mathematica's function ListConvolve that implicitly uses many processor
cores. For this reason, no explicit parallelization was done in our
program. 

\begin{figure}[ht]
\vspace{0.5cm}
\centering{}\includegraphics[width=8cm]{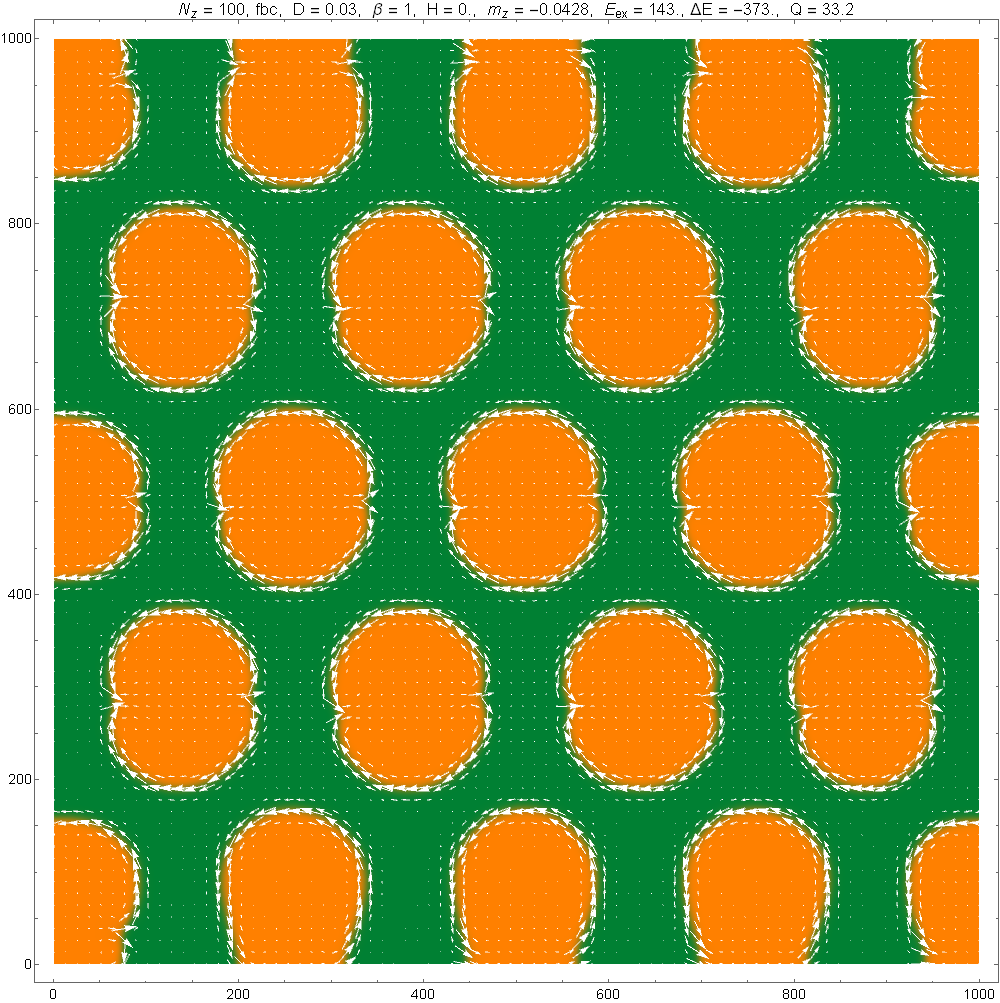}
\caption{Biskyrmion bubble lattice with lower energy, $\Delta E = - 373$, that evolves from the Weierstrass lattice of longer period under the action of PMA and DDI at $H=0$, $\beta=1$, and $D = 0.03$.}
\label{Fig-large-bubbles} 
\end{figure}

For each skyrmion state we compute the topological charge $Q$, the exchange energy $E_{ex}$, the total energy $E$, and the difference, $\Delta E$, of the total energy from the energy of the uniformly magnetized state. The difference of the exchange energy from $4\pi JQ$ is the measure of the distortion of BP biskyrmions, while the sign of $\Delta E$ is indicative of whether the formation of the biskyrmion lattice can lower the total energy. Our numerical results for the energies are presented in the units of $J$.

Typical results of the energy minimization at $H=0$ and $\beta=1$ for two values of PMA, starting with the same Weierstrass
biskyrmion lattice as the initial condition,  are illustrated by Figs.\  \ref{Fig-biskyrmion-lattice} and \ref{Fig-bubble-lattice}.  The final state does not depend on the separation $d$ in the biskyrmions but depends on the lattice period which defines the total topological charge of the system. The latter is approximately conserved and changes only slightly due to the use of the open boundaries. At low PMA, $D/J=0.001$, the system relaxes to a biskyrmion lattice shown in Fig. \ref{Fig-biskyrmion-lattice}. Its energy, however, is above the energy of the uniformly magnetized state, $\Delta E > 0$, indicating that it can only be a metastable state stabilized by the conservation of the topological charge. 

At $D/J=0.03$ the initial Weierstrass biskyrmion lattice evolves into a lattice of biskyrmion bubbles shown Fig.\ \ref{Fig-bubble-lattice}. The energy of such a lattice is below the energy of the uniformly magnetized state, $\Delta E < 0$, indicating that it can be a stable domain structure at $H = 0$. It consists of regularly spaced magnetic bubbles with $Q=2$, each having two Bloch lines. Notice that Bloch lines in biskyrmion bubbles forming a triangular lattice at $H = 0$ have been observed in experiment \cite{Yu-2014}.

Even a lower energy can be achieved by increasing the period of the lattice, which leads to bigger biskyrmion bubbles in the final state, see Fig.\ \ref{Fig-large-bubbles}. This finding confirms conjecture of Ref.\ \onlinecite{Loudon2019} that the absolute minimum of the energy is likely to correspond to the limiting case of biskyrmion bubbles of size much greater than the domain wall thickness.

\section{Discussion} \label{Sec_Discussion}
We have studied biskyrmion lattices in non-chiral magnetic films of finite thickness. While such lattices have been observed in experiments, their theoretical description presented a considerable challenge. Our approach has been based upon observation that representation of ferromagnetically-coupled spins in the complex plane, $z = x + iy$, that uses a complex function $\omega(z) = (s_x + is_y)/(1-s_z)$, naturally provides biskyrmion lattices when $\omega(z)$ is written in terms of elliptic functions. Starting with such a function as an initial condition, we study evolution of the spin configuration under the action of perpendicular magnetic anisotropy (PMA) and dipole-dipole interaction (DDI).

Our main result is that in a certain a range of parameters, that is relevant to experimental situations, the energy of a triangular lattice of biskyrmion bubbles is significantly lower than the energy of the uniformly magnetized film in a zero magnetic field. Notice in this connection that for a large, highly non-linear, hysteretic magnetic system that has infinite number of local energy minima one cannot find and compare energies of all metastable non-uniformly magnetized states that emerge due to the exchange interaction, PMA, and DDI (such as various kinds of labyrinth domains, various kinds of cylindrical domains or skyrmions, with and without Bloch lines and Bloch points, etc.). Consequently, it is impossible to say with the absolute certainty that a biskyrmion lattice observed in experiments and simulations is the ground state of the system. Its low energy computed in our model, however, makes it plausible that it represents a kind of the energy-minimizing domain structure that has not been observed until recently. 

When biskyrmion lattices correspond to the energy minimum, they are likely to consist of bubbles of size much greater than the domain wall thickness. An additional connection between theory and observation is provided by the fact that biskyrmions arranged in a triangular lattice possess Bloch lines in both, experiments and computations. Further experimental studies of magnetic phases of non-chiral films along the hysteresis loop will shed more light on the place of biskyrmion lattices in the phase diagram. 

\section{Acknowledgements}

This work has been supported by the grant No. DE-FG02-93ER45487 funded
by the U.S. Department of Energy, Office of Science.

\end{document}